\documentclass{aa}  

\usepackage{afterpage,natbib,lipsum,ulem}
\usepackage{graphicx}
\usepackage[usenames, dvipsnames]{color}

\newcommand{\lya}{Ly-$\alpha$ }

\definecolor{darkgreen}{rgb}{0.1, 0.45, 0.15}

\usepackage{txfonts} 
\begin{document}

   \title{A hierarchical field-level inference approach to reconstruction from sparse Lyman-$\alpha$ forest data}

   \author{Natalia Porqueres\inst{1} \and Oliver Hahn\inst{2} \and Jens Jasche\inst{3} \and Guilhem Lavaux\inst{4}} 

   \institute{ Imperial Centre for Inference and Cosmology, Imperial College London, Blackett Laboratory, Prince Consort Road, London SW7 2AZ, United Kingdom
         \and
         Laboratoire Lagrange, Universit\'{e} C\^{o}te d'Azur, Observatoire de la C\^{o}te d’Azur,
CNRS, Blvd de l’Observatoire, CS 34229, 06304 Nice, France 
         \and
         The Oskar Klein Centre, Department of Physics, Stockholm University, Albanova University Center, SE 106 91 Stockholm, Sweden
         \and
         CNRS \& Sorbonne Universit\'{e}, UMR7095, Institut d'Astrophysique de Paris, F-75014, Paris, France
         }

   \date{Received 23/05/2020; accepted 14/08/2020}

 
  \abstract
  %
   {We address the problem of inferring the three-dimensional matter distribution from a sparse set of one-dimensional quasar absorption spectra of the Lyman-$\alpha$ forest. Using a Bayesian forward modelling approach, we focus on extending the dynamical model to a fully self-consistent hierarchical field-level prediction of redshift-space quasar absorption sightlines. Our field-level approach rests on a recently developed semiclassical analogue to Lagrangian perturbation theory (LPT), which improves over noise problems and interpolation requirements of LPT. It furthermore allows for a manifestly conservative mapping of the optical depth to redshift space. In addition, this new dynamical model naturally introduces a coarse-graining scale, which we exploited to accelerate the Markov chain Monte-Carlo (MCMC) sampler using simulated annealing. By gradually reducing the effective temperature of the forward model, we were able to allow it to first converge on large spatial scales before the sampler became sensitive to the increasingly larger space of smaller scales. We demonstrate the advantages, in terms of speed and noise properties, of this field-level approach over using LPT as a forward model, and, using mock data, we validated its performance to reconstruct three-dimensional primordial perturbations and matter distribution from sparse quasar sightlines.}


   \titlerunning{A hierarchical field-level inference approach to reconstruction from Ly-$\alpha$ forest data}
   \authorrunning{Porqueres et al.}
    
   \maketitle
  
%

\section{Introduction}
    
    A fundamental task in cosmology consists of relating the structures we see in the late Universe with the primordial density fluctuations when the Universe was still close to homogeneous. While the primordial density fluctuations are well-described by a Gaussian distribution \citep{Planck2018nonGaussianity}, gravitational collapse in the cosmological context leads to intricate structures with a large density contrast over the age of the Universe. Non-linear dynamics produce a matter density field with a highly non-Gaussian complex statistical structure, which makes the analysis of the late-time Universe very challenging. A detailed modelling of the cosmic matter distribution would require describing the high-order statistics corresponding to the filamentary structure of the cosmic web. At present, a closed-form description of the non-linear density field in terms of a high-dimensional multivariate probability distribution does not exist. Although there are approximations to reproduce the statistical behaviour of the dark matter density field (e.g. log-normal distribution or multivariate Gaussians, \citealt{lahav1994wiener, zaroubi1999wiener, kitaura2008bayesian, kitaura2009cosmic,  HADES}), they only parameterise the one- and two-point statistics and fail to reproduce more complex structures such as filaments \citep{DensityLikelihood95, DensityLikelihood96, DensityLikelihood03}.

    Here, we instead follow the forward modelling approach to relate initial conditions and observables. It consists of a data model that describes how the continuous three-dimensional field of initial matter fluctuations affects a set of predicted observables, which are then compared to data.  In this work, we specifically focus on the observed flux in quasar absorption spectra as our observable. We are then interested in the inverse problem: given a set of quasar spectra, we want to infer the underlying matter distribution and the corresponding primordial fluctuations. In this context, the data model should describe everything that may happen between the initial fluctuations and the observation of the spectra, which includes the time-evolution of the matter density and the cosmic structure formation model, as well as sparse sampling of the observable. In this way, every data point is used, rather than relying on summary statistics that do not capture all the information and whose distributions are not well known.

    Structure formation in $\Lambda$ cold dark matter ($\Lambda$CDM) cosmology proceeds through the collapse of baryons and dark matter that can be well approximated as cold in comparison to the velocities induced by gravity. The dynamics and growth of cosmic structures are then described by Lagrangian perturbation theory \citep[LPT; ][]{Zeldovich:1970, Bouchet:1992} well, which directly describes the motion of fluid elements. LPT is, however, only valid before the crossing of fluid trajectories and, therefore, restricted to large scales or early times. Due to its simplicity, especially when truncated at first or second order, many forward modelling approaches in cosmology rely on LPT to describe the dynamics of (cold collisionless) matter \citep[see, e.g.][]{BORG,KitauraLPT,WangLPT,BosLPT,AtaLPT}. Since it is no longer valid at and after shell-crossing, LPT has to be employed in a way that shell-crossed scales are filtered out prior to employing it \citep[see e.g.][]{Sahni1995}. However, determining the scale of shell-crossing is approximate and usually challenging.
    
    Another downside of LPT is that it predicts fluid trajectories, while in many cases one is instead interested in density or velocity fields at fixed spatial, that is Eulerian, coordinates. Numerically, an Eulerian density field can only be obtained by interpolating the fluid elements back to an Eulerian grid. This can be achieved using a particle injection scheme such as cloud-in-cell deposit \citep[CIC;][]{Hockney:1981}, which is impacted by particle sampling noise. This can also be achieved by interpolating from a tessellation of the distribution function \citep[e.g.][]{Abel:2012}, which is computationally more expensive than CIC. Using Eulerian perturbation theory is not an option since it requires going to very high orders to achieve comparable accuracy to LPT \citep[e.g.][]{Bouchet96}.
    
    An alternative exists in `field-based' approaches that directly operate by predicting the Eulerian density based on the notion of particle trajectories. The semiclassical approach to describe cold collisionless dynamics presented in \cite{Ulhemann19}, but see also \cite{Short:2006a,Short:2006b}, provides such an alternative to LPT. This approach, which we name as propagator perturbation theory (PPT) here, translates LPT into an action and then uses a propagator to evolve a wave function, which encodes the cosmological perturbations. From the evolved wave function, the Eulerian density and velocity fields are readily obtained. The PPT approach introduces an additional free parameter, an effective $\hbar$, which acts as a natural smoothing, or coarse-graining scale. We note that PPT is fundamentally different from Schroedinger-Poisson (SP) analogues as effective models \citep[cf.][]{Widrow1993,Uhlemann2014,Kopp2017,Garny:2020,Eberhardt2020} of cold Vlasov-Poisson (VP) dynamics (VP underlies all CDM non-linear cosmological structure formation; \citealt{Peebles1980}). PPT is not a fully non-linear model like SP but a perturbative analogue to LPT, more similar in spirit to the Burgers approach of \cite{Matarrese2002}. However, PPT gives easier access to phase space statistics than LPT by absorbing the `sum-over-streams' into a propagator \citep[cf.][]{Ulhemann19}.
    
    Before shell-crossing, the first order of the PPT approach provides results equivalent to the first-order LPT in the limit of vanishing $\hbar$. At shell-crossing, while LPT leads to infinite densities, the PPT density remains finite and, after shell-crossing, the PPT density presents interference patterns in multi-stream regions. These interference patterns, therefore, provide a natural way to detect the shell-crossing scale. These oscillations naturally encode stream-averaged velocity fields (and higher moments) that are notoriously expensive to obtain for cold Vlasov dynamics \citep[cf.][]{Pueblas2009,Hahn2015,Buehlmann2019}. 
    
\begin{figure}[t]
        \centering
        \includegraphics[width=\hsize,clip=true]{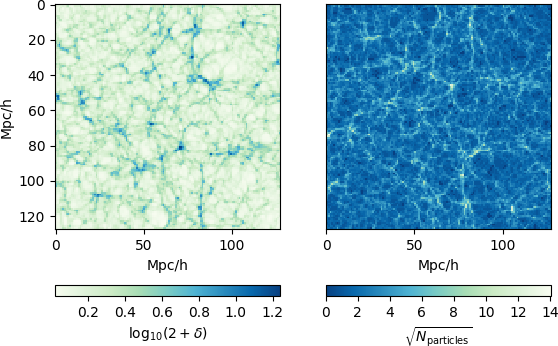}
        \caption{Density field obtained with LPT (left) and corresponding signal-to-noise due to the particle distribution (right). In the LPT model, particles cluster at high-densities, poorly sampling the low-density regimes from which the Ly-$\alpha$ forest arises. }\label{fig:noise_lpt}
    \end{figure}

    By predicting Eulerian fields, PPT overcomes the particle sampling problem of LPT where particles cluster in the high-density regions and the under-densities are affected by high levels of shot noise, as illustrated in Fig.~\ref{fig:noise_lpt}. This is especially relevant for the analysis of Lyman-$\alpha$ (Ly-$\alpha$) forest observations since these data are particularly sensitive to under-dense regions in the matter distribution \citep[cf.][]{Peirani14, Sorini16}.
    
    At present, major analyses of the \lya forest focus only on the analysis of the matter power spectrum \citep[e.g.][]{Croft98,CosmoLyaSeljak,CosmoLyaViel,CosmoLyaSeljak,Bird11,CosmoParamsSlosar,CosmoParamsBusca,NeutrinoLya15,NeutrinoLya15b, ReionLyaNasir, LyaNeutrinos17,LyaNeutrinos17Rossi,CosmoParamsLyaBautista,ReionLyaBoera,BAOeBOSSLya,Maitra2019}. However, these approaches ignore significant amounts of information contained in the higher-order statistics of the matter density field as generated by non-linear gravitational dynamics in the late time universe \citep[][]{cmb_lensing_filaments}. While various approaches to perform three-dimensional density reconstructions have been proposed in the literature, they are based on Wiener filter techniques \citep{Ozbek16,Stark15,WienerRavaoux,LATIS}  or they assume the density amplitudes to be log-normally distributed \citep{Kitaura12,Gallerani11}. These approaches fail to reproduce the high-order statistics of the filamentary matter distribution. To reproduce the high-order statistics, \cite{PorqueresLya} and \cite{Horowitz19} recently used a large-scale optimisation approach to fit a gravitational structure growth model to simulated Ly-$\alpha$ data.

    In this work, we employed for the first time PPT in a Bayesian forward model to infer the dark matter density from the \lya forest. In particular, we use the extension of the BORG framework \citep{HADES, BORG, BORG-3} to the analysis of the \lya forest presented in \cite{PorqueresLya}, combined with a redshift-space optical depth field obtained from our extension of PPT presented here.
    
    Our inference framework consists of a Gaussian prior on the primordial matter fluctuations, a physical model of structure formation to evolve the density field (in this case, the PPT), and a likelihood based on the fluctuating Gunn-Peterson approximation \citep[FGPA,][]{GPeffect}. To extract the large-scale structure information from the data, the BORG framework employs a Markov chain Monte-Carlo (MCMC) sampler. MCMC methods typically require a warm-up phase before they reach the target distribution and acquire a stationary state. This warm-up phase can be computationally expensive. 
    
    To accelerate the warm-up phase, we exploit the fact that PPT comes with a built-in tuneable scale parameter, $\hbar$, which controls an effective phase space resolution. One can think of this as an effective temperature so that a large $\hbar$ corresponds to a high temperature. The effective temperature controls to which features particle trajectories will be able to respond.
    In this work, we show that the computational costs of the warm-up phase can be reduced by performing a simulated annealing with the PPT model and taking advantage of the lower complexity of coarser scales. Such a procedure has been explored in the field of image processing \citep{Annealing89, Annealing03} and consists of walking down a hierarchy of scales from coarsest to finest resolution. At one level of resolution, we can focus on a particular scale: coarsest scales are frozen from the lowest resolution, and smallest scales are still evolving and will continue fluctuating in higher resolutions. 
    Decreasing the effective $\hbar$ over the course of the chain thus corresponds to annealing and allows the trajectories to respond to increasingly finer structures. For high-$\hbar$ modelling, low spatial resolution can be used, allowing for further speed-up. By consistently changing the resolution and $\hbar$, we can then perform a simulated annealing that substantially reduces the computational cost of the warm-up phase of the MCMC sampler. 
    
    The paper is organised as follows. Section \ref{sec:qlpt} provides a brief description of the PPT model and its extension to include redshift space distortions. Section \ref{sec:borg} gives an overview of our Bayesian inference framework, BORG, as required for this work. In Section \ref{sec:data}, we described the simulated data employed in testing and validating the method. The simulated annealing is described in Section \ref{sec:annealing}, showing that this strategy reduces the computational cost of the warm-up phase of the Markov sampler. The inference results from \lya forest data in redshift space are presented in Section \ref{sec:run_qlpt}. Finally, Section \ref{sec:conclusions} summarises the results. 
    
\section{Propagator perturbation theory for the \lya forest} \label{sec:qlpt}
    In this section, we briefly describe the PPT model as relevant for this work and present its extension to redshift space. For a more detailed description, its derivation and proof that it has LPT as its classical limit, we kindly refer the reader to \cite{Ulhemann19}.
    
    \subsection{Background}
    \label{sec:QLPT_background}
    The PPT model relies on semiclassical dynamics to evolve the dark matter density using a propagator. In the Zel'dovich approximation, a fluid element moves in a time $D_+$ from its initial (Lagrangian) coordinate $\vec{q}$ to its final (Eulerian) coordinate $\vec{x}$ on a straight line, where $D_+$ is the linear theory growth factor. The classical action of this motion is therefore simply
    \begin{equation}
        S_0(\vec{x},\vec{q};a) = \frac{1}{2}\frac{(\vec{x}-\vec{q})^2}{D_+(a)}.
    \end{equation}
    This action can be promoted to a (free) propagator $K_0$ using the Dirac-Feynman \citep{Dirac1933,Feynman:1948} trick
    \begin{equation}
        K_0(\vec{x},\vec{q};a) = (2\pi i \hbar D_+(a))^{-3/2} \, \exp\left[\frac{i}{\hbar}S_0(\vec{x},\vec{q}; a)\right].
    \end{equation}
    This propagator can be used to compute the transition amplitude from an initial state represented by a wave function $\psi_0$ to the final state
    \begin{equation}
    \psi(\vec{x},a) = \int {\rm d}^3q\,K_0(\vec{x},\vec{q};a)\,\psi_0(q).\label{eq:propagation}
    \end{equation}
    Herein, $\hbar$ has no physical meaning, but instead is a free parameter that controls an effective smoothing scale, which we exploited to our advantage later. \cite{Ulhemann19} have shown that this approach converges rigorously to the Zel'dovich approximation, and can be upgraded to second order LPT by adding a time-independent potential to the action. Here we only considered the free propagator however.

    \begin{figure*}[t]
        \centering
        \includegraphics[width=\hsize,clip=true]{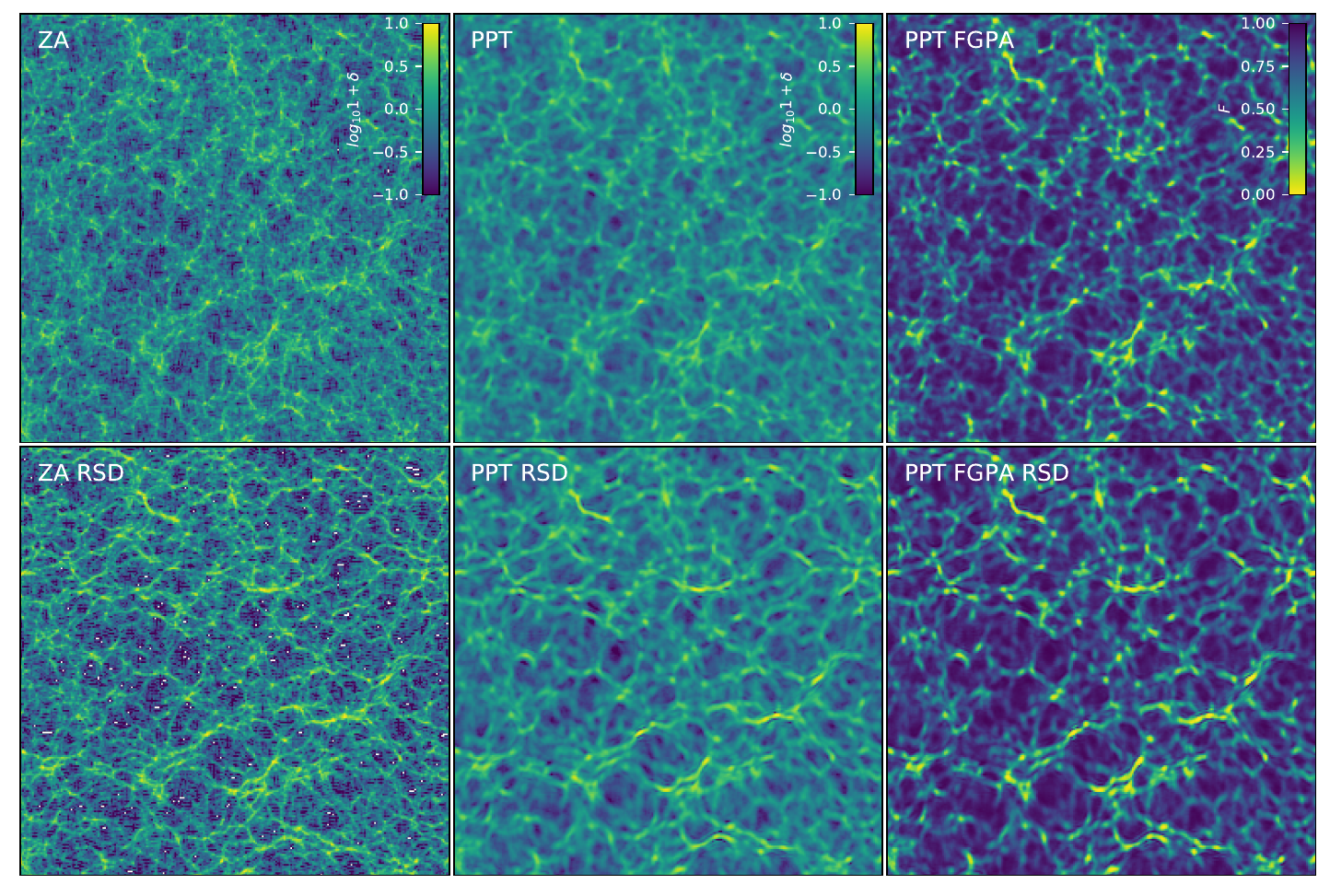}
        \caption{Density fields (left and middle column) and quasar flux field in the FGP approximation (right column) in physical space (top panels) and in redshift space (bottom panels), with the line-of-sight direction upwards. The leftmost panels use the Zel'dovich approximation and CIC deposit, the others use the PPT formalism. In all cases we used $256^3$ resolution elements, the box size (and extent of each image) used for this comparison is $256\,h^{-1}{\rm Mpc}$ at $z=2.5$. The thickness of the projected slice is $2h^{-1}{\rm Mpc}$ and white pixels in the CIC panels indicate zero particles deposited.}\label{fig:qlpt_rsd}
    \end{figure*}
    
    \subsection{Implementation} \label{sec:implementation}
    
    We now describe our specific implementation of PPT and how late-time density and velocity fields can be obtained. Since the Zel'dovich approximation has pure growing-mode solutions only, the initial state has only one degree of freedom, the `back-scaled'\footnote{Back-scaling means that we think of the potential at the initial time $\phi^{\rm ic}$ and the linear potential $\phi$ at the target time $a_{\rm target}$ related by the factor $\phi^{\rm ic} = \phi\times \frac{a_{\rm target}}{D_+(a_{\rm target})}\lim_{a\to0} \frac{D_+(a)}{a}$.} gravitational potential $\phi^{\rm ic}$ which is given at a fictitious initial time when $D_+=0$ and represents a homogeneous state of the Universe before structure formation begins. The corresponding wave function has to have only phase perturbations and is given by
    \begin{equation}
    \psi_0(\vec{q}) := \exp\left[ -\frac{i}{\hbar} \phi^{\rm ic}(\vec{q})\right].
    \label{eq:psi0}
    \end{equation}
    We note that the density associated with the initial wave function $\rho_0:=\psi_0 \overline{\psi}_0 = 1$ corresponds to the uniform mean density (an overline denotes a complex conjugate). We often refer also to the primordial density fluctuations, by which we mean the field $\delta^{\rm ic}:=\nabla^2\phi^{\rm ic}$, corresponding to the linear theory total matter density with the growth factor scaled out. 
    
    The evolved state $\psi$ is obtained by computing the propagation of the field $\psi_0$ through eq.~(\ref{eq:propagation}), which mathematically corresponds to a convolution integral. This is numerically most conveniently carried out as a multiplication in Fourier space. Using the discrete Fourier transform (DFT), and exploiting circular convolution on a periodic domain, we find the full expression relating $\phi^{\rm ic}(\vec{q)}$ and $\psi$ as
    \begin{equation}
        \psi(\vec{x},a) = {\rm DFT}^{-1}\left[ \exp\left(-i\hbar \frac{k^2}{2}D_+(a)\right) \,{\rm DFT}\left[\exp\left(-\frac{i}{\hbar}\phi^{\rm ic}(\vec{q})\right)\right]\right],
    \label{eq:psi}
    \end{equation}
    where $\vec{q}$ and $\vec{x}$ are discrete on a regular three-dimensional grid.

    By construction, this wave function encodes all the phase-space information, that is, the full cumulant hierarchy \citep[cf.][]{2018JCAP...10..030U}. In our case we are interested in the normalised density $\rho := 1+\delta$, where $\delta$ is the fractional overdensity, and the peculiar momentum field  $\vec{j} := (1+\delta) \vec{v}$, where $\vec{v}$ is the peculiar velocity. These are given in terms of the propagated wave function $\psi=\psi(\vec{x},a)$ as
    \begin{equation}
     \rho=\psi\overline{\psi}\qquad\textrm{and}\qquad  \vec{j} = \frac{i\hbar}{2} \left( \psi \nabla \overline{\psi} - \overline{\psi} \nabla \psi \right).
    \label{eq:rho}
    \end{equation}
    This density agrees with a smoothed version of the Zel'dovich approximation before shell-crossing, where $\hbar$ controls the smoothing \footnote{Following from Nyquist-Shannon, the smallest possible $\hbar$ fulfills the condition $|\Delta \phi|\ / \ \hbar < \pi$, where $\Delta \phi$ is the difference of the gravitational potential in neighbouring voxels.}  \citep{Short:2006a,Short:2006b,Ulhemann19}. After shell-crossing, the Zel'dovich approximation is of course no longer valid since it does not account for secondary infall, and collapsed structures simply disperse again. More importantly, shell-crossing is accompanied by the formation of caustics, regions of infinite density, and multi-stream flow \citep[cf.][]{Arnold:1982,Hidding:2014}. While the density in the classical approach becomes infinite or multi-valued, in PPT $\rho$ remains finite and develops interference patterns in multi-stream regions, all regulated by the finite $\hbar$.

    
\subsection{Redshift-space distortions} \label{sec:rsd_propagator}
Cosmological observations take place in redshift space rather than in comoving physical space. For all practical purposes, we can make the approximation of a distant observer, which implies that the redshift space distortion can be chosen to coincide with a Cartesian axis. Specifically, a particle is not observed to be at its Eulerian position $\vec{x}$, as we discussed in Sec.~\ref{sec:QLPT_background}, but instead at its redshift space position $\vec{s}$, because of deviations from pure Hubble expansion (peculiar velocities). In LPT, this is given by 
\begin{equation}
    \vec{s} := \vec{x} + f(a)\, \left(\vec{\Psi}\cdot \hat{\vec{e}}_{\rm LOS}\right)\hat{\vec{e}}_{\rm LOS},
\end{equation}
where $\hat{\vec{e}}_{\rm LOS}$ is a unit vector pointing along the line-of-sight (which we shall without loss of generality assume to be along the z-axis), $\vec{\Psi}$ is the displacement field between Lagrangian and Eulerian coordinates, $\vec{\Psi}:=\vec{x}-\vec{q}$, and $f = {\rm d}\log D_+ / {\rm d}\log a$. This is quite obviously simply a velocity dependent displacement, and it can therefore be trivially included in an additional propagator from Eulerian to redshift space, given as
\begin{equation}
    K_{\rm RSD}(\vec{s},\vec{x};a) = N \, \exp\left[ \frac{\rm i}{\hbar} \frac{1}{2} \frac{\left( \left(\vec{s}-\vec{x}\right)\cdot \hat{\vec{e}}_{\rm LOS} \right)^2}{f(a)\,D_+(a)}\right],  \label{eq:rsd_prop}
\end{equation}
with $N$ a normalisation that has to be suitably chosen. Effectively, at leading order PPT, the propagators can be trivially combined into a single propagator from Lagrangian space to redshift space, which in Fourier space takes the form
\begin{equation}
    \widehat{K}(\vec{k};\,a) := \exp\left[ -\frac{{\rm i}\hbar}{2} \left(k^2 + f(a)\, \left(\vec{k}\cdot\hat{\vec{e}}_{\rm LOS}\right)^2 \right)\,D_+(a) \right].
    \label{eq:propagator_rsd}
\end{equation}
While we did not use the next-to-leading order (NLO) version of PPT \citep[cf.][]{Ulhemann19} here, the propagation to redshift space can also be applied at NLO, by performing the propagation to redshift space after carrying out the `kick-drift-kick' endpoint approximation to the path integral (their eq.~D4).

\subsection{Modelling of the \lya-forest} \label{sec:qlpt_rsd_lya}
We have explained above how PPT can be used to predict a quasi-linear density field $\rho=\psi\overline{\psi}$, consistent with the Zel'dovich approximation, from a wave function $\psi$ propagated forwards to time $a$. In order to model the absorption of photons from the quasar, we employed the fluctuating Gunn-Peterson approximation \citep{GPeffect}. The fractional transmitted flux is given by
\begin{equation}
    F = e^{-\tau},
\label{eq:fgpa_flux}
\end{equation}
where $\tau$ is the optical depth. In Eulerian space, the optical depth field reads
\begin{equation}
    \tau(\vec{x}) := A\,\rho(\vec{x})^\beta,
\label{eq:fgpa_tau}
\end{equation}
where $A$ and $\beta$ are heuristic parameters, which are given by the physical state of the intergalactic medium. In a next and final step, we want to map this optical depth to redshift space and compute the transmitted quasar flux. In order to achieve this, we construct a new wave function that transports the optical depth, that is, we re-scale the amplitude such that
\begin{equation}
    \chi_0(\vec{x}) := \sqrt{A}\,\rho^{\frac{\beta-1}{2}}(\vec{x}) \psi(\vec{x}).
\end{equation}
This new wave function obeys $\chi_0\overline{\chi}_0=\tau$, while the phase information (i.e. the velocity) of the evolved wave function $\psi$ is untouched. We note that this is essentially just a direct application of Madelung's interpretation of the wave function \citep{Madelung:1927}. This means we can exploit that the amplitude of the wave function has a conserved current. This is a crucial property since the optical depth $\tau$ is conserved under the mapping between physical and redshift space \citep[e.g.][]{Seljak12}. We can therefore use the RSD propagator in eq.~(\ref{eq:rsd_prop}) to propagate the $\chi$ field to redshift space, manifestly conserving $\tau$, by evaluating
\begin{equation}
    \chi(\vec{s}) := \int {\rm d}^3x\,K_{\rm RSD}(\vec{s},\vec{x};a)\,\chi_0(\vec{x}).
    \label{eq:chi}
\end{equation}
We note that this is essentially a `non-linear velocity RSD' \citep[cf.][their eq. 4.3, but at a field level and using quasilinear Zel'dovich velocities at the order we are considering here]{Cieplak:2016}. In a final step, we can obtain the three-dimensional quasar flux field $F$ in redshift space by evaluating
\begin{equation}
    F(\vec{s}) = \exp\left[ - \chi\overline{\chi}\right].
\end{equation}
    
In Fig.~\ref{fig:qlpt_rsd}, we show a comparison of the density field obtained with PPT (right panels) and the Zeldovich approximation (left panels) in comoving physical space (top panels), and in redshift space (bottom panels), showing that PPT and LPT provide the same structures in real and redshift space. In the right panels, we show the quasar flux field $F$ in physical and in redshift space. 
  
   \begin{figure}
        \centering
            {\includegraphics[width=0.9\hsize,clip=true]{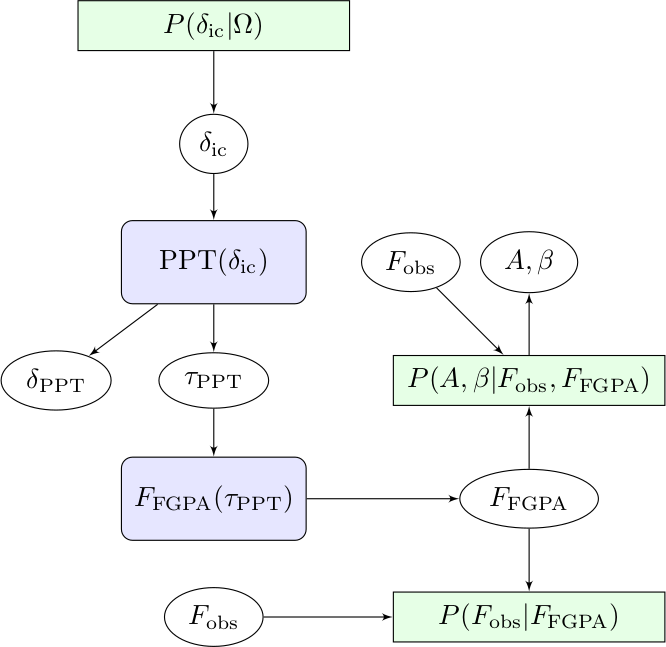}}
        \caption{Hierarchical representation of the BORG inference framework for the analysis of \lya forest data. Primordial fluctuations $\delta_\mathrm{ic}$ encoded in a a set of Fourier modes at $z\approx 1000$ are obtained from the prior $P(\delta_\mathrm{ic}|\Omega)$, where $\Omega$ represents the cosmological parameters. These initial conditions are evolved to $z=2.5$ using PPT, which provides the optical depth $\tau_\mathrm{PPT}$ and the evolved density, $\delta_\mathrm{PPT}$.  The optical depth is then used to generate quasar spectra based on the fluctuating Gunn-Peterson approximation (FGPA). $F_\mathrm{obs}$ indicates the data. Purple boxes indicate deterministic transition while green boxes are probability distributions. Iterating this procedure results in a Monte Carlo Markov Chain that explores the joint posterior distribution of the three-dimensional matter distribution underlying \lya forest observations.} 
        \label{fig:borg}
    \end{figure}
    
    \begin{figure*}[t]
        \centering
        \includegraphics[width=0.8\hsize,clip=true]{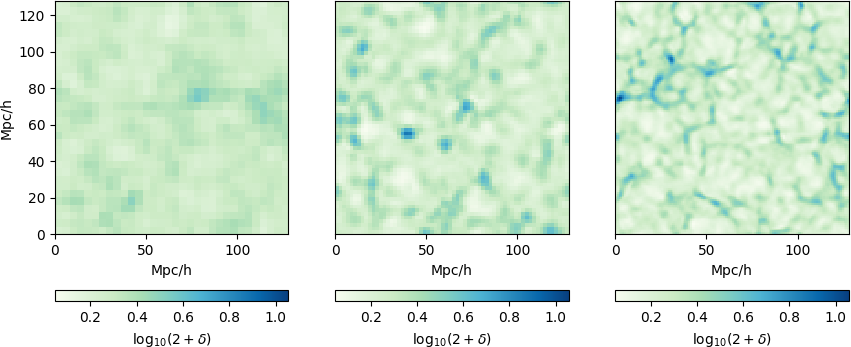}
        \caption{Annealing of the density field. The density field is inferred hierarchically, starting with the largest scales and, once these are converged, opening new modes. These panels correspond to $N=32^3$ voxels (left),  $N=64^3$ (middle) and $N=128^3$ (right). The $\hbar$ parameter is decreased over the course of the chain, allowing the algorithm to respond to finer structures. From left to right, $\hbar$ decreased from 0.15 to 0.09. }\label{fig:annealing_density} 
    \end{figure*}
    
\section{The BORG framework for \lya forest} \label{sec:borg}

    As mentioned above, we implemented PPT as a forward model in the BORG framework to infer the three-dimensional matter distribution underlying \lya forest data. In this section, we provide a summary of the algorithm. A more detailed description of the BORG framework can be found in \cite{BORG, JLW15, Lavaux16, Jasche18} and, more specifically, the extension of BORG to the \lya forest analysis is described in \cite{PorqueresLya}. 
    
    The BORG framework is a Bayesian inference method aiming at inferring the non-linear spatial dark matter distribution and its dynamics from cosmological data sets. The underlying idea is to fit full dynamical gravitational and structure formation models to observations. By using non-linear structure growth models, the BORG algorithm can exploit the full statistical power of high-order statistics of the matter distribution imprinted by gravitational clustering. This dynamical model links the primordial density fluctuations to the present large-scale structures. Therefore, the forward modelling approach allows translating the problem of inferring non-linear matter density fields into the inference of the spatial distribution of the primordial density fluctuations, which are well described by Gaussian statistics \citep{Planck2018nonGaussianity}. The BORG algorithm, therefore, infers the initial matter fluctuations, the dark matter distribution and its dynamical properties from observations. 

    While the BORG framework incorporates several dynamical models based on Lagrangian perturbation theory and particle-mesh models, in this work, we include the PPT. Besides the advantage to work directly with fields, PPT allows for a reduction of the computational costs of the algorithm (see Section \ref{sec:annealing}). 
    
    We tested the inference with PPT by applying the BORG framework to the analysis of simulated \lya forest data. The field-based approach of the PPT provides an advantage when analysing \lya forest observations since these data arise from low-density regimes, which are impacted by particle sampling noise in the standard LPT. To model the \lya forest, we used a Gaussian likelihood based on the fluctuating Gunn-Peterson approximation,
    \begin{align}
        P(\delta^\mathrm{ic},\delta^\mathrm{f}|F) =  \prod_{n,x} \frac{1}{\sqrt{2\pi\sigma^2}}  \exp\Bigg[-\frac{\Big((F_n)_x - \exp(-\tau)\Big)^2}{2\sigma^2}\Bigg],
        \label{eq:likelihood_lya}
    \end{align}
    where $n$ labels the lines of sight and $x$ runs over the pixels along a line of sight.
    The hierarchical representation of the algorithm is illustrated in Fig.~\ref{fig:borg}.
    
    At its core, the BORG framework employs MCMC techniques. This method allows inference of the full posterior distribution from which we can quantify the uncertainties in our results. However, the inference of the density field typically involves $\mathcal{O}(10^7)$ free parameters, corresponding to the discretised volume elements of the observed domain. To explore efficiently this high-dimensional parameter-space, the BORG framework uses a Hamiltonian Monte Carlo (HMC) method, which exploits the information in the gradients and adapts to the geometry of the problem. We need, therefore, the gradient of the dynamical forward model. The PPT gradient is derived in Appendix \ref{app:qlpt_gradient}. More details about the HMC and its implementation are described in \cite{jasche2010fast} and \cite{jasche2013bayesian}.

\section{The data} \label{sec:data}
    To test the inference framework, we generated artificial mock observations emulating the properties of present \lya forest surveys such as the CLAMATO survey \citep{ClamatoMock, ClamatoDR1} and LATIS \citep{LATIS}. In this section, we describe the properties of the artificial data.  
   
    Mock data are constructed by first generating Gaussian initial conditions on a cubic Cartesian grid of side length of 128$h^{-1}$~Mpc with a resolution of 1$h^{-1}$~Mpc. To generate primordial Gaussian density fluctuations we used a cosmological matter power-spectrum including the baryonic wiggles calculated according to the prescription provided by \cite[][]{EH98,EH99}. We further assumed a standard $\Lambda$CDM cosmology with the following set of parameters: $\Omega_m = 0.31,\ \Omega_\Lambda = 0.69,\ \Omega_b = 0.022,\ h=0.6777,\ \sigma_8= 0.83,\ n_s = 0.9611$ \citep{Planck15}. Here ${\rm H}_0=100 h$~km~s$^{-1}$~Mpc$^{-1}$. 
    
    To generate realisations of the non-linear density field, we evolve the Gaussian primordial fluctuations via the forward model (PPT or LPT, respectively). A three-dimensional quasar flux field is generated by applying the FGPA model in eq. \ref{eq:fgpa_flux} and \ref{eq:fgpa_tau}, assuming constant parameters $A = 0.35$ and $\beta=1.56$ at $z=2.5$, corresponding to the values in \cite{ClamatoMock}. From this three-dimensional quasar flux field, we generate individually observed skewers by tracing lines of sight through the volume. Specifically, we generate a total of $1024$ lines of sight parallel to the $z$-axis of the box, randomly distributed with a mean separation of  8 h$^{-1}$ Mpc.  The separation between lines of sight is the most important parameter of Ly-$\alpha$ surveys. Present surveys like CLAMATO \citep{ClamatoDR1} and LATIS \citep{LATIS} achieve an average separation of 2.4 h$^{-1}$ Mpc. Finally, we added Gaussian pixel-noise to the flux with $\sigma = 0.03$. This $\sigma$ results in a signal-to-noise of $\mathrm{S/N}=2$, which corresponds to the majority of lines of sight in the CLAMATO survey. 
    
\section{Simulated annealing} \label{sec:annealing}
    In this section, we describe how we use a simulated annealing strategy to accelerate the warm-up phase of the MCMC sampler. Specifically, PPT has a built-in coarse-graining scale through the effective $\hbar$ which one can think of as an effective energy (or temperature) scale. By gradually reducing $\hbar$, one can, therefore, implement a simulated annealing procedure for the forward model. We compare the computational costs of the warm-up phase to the standard LPT approach. In this comparison, we use the PPT in physical (not redshift) space, as described in Section~\ref{sec:implementation}.

    \begin{figure*}
        \centering
        \begin{tabular}{c c}
            \includegraphics[width=0.45\hsize,clip=true]{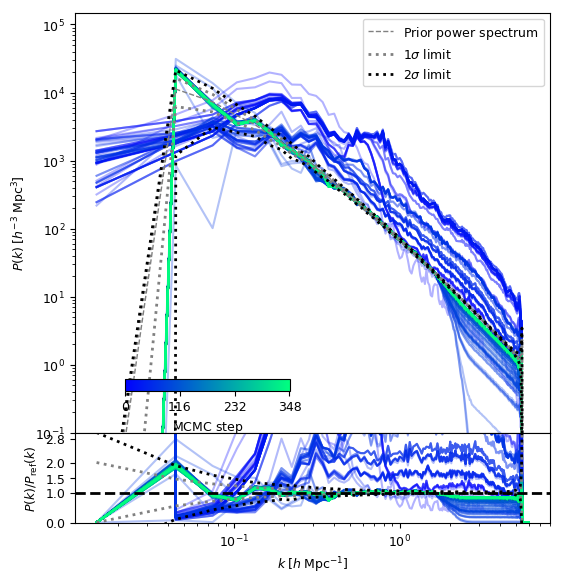} & \includegraphics[width=0.45\hsize,clip=true]{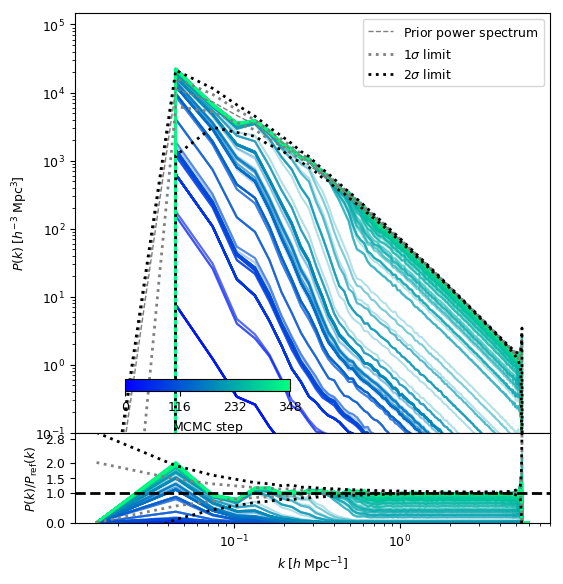} \leavevmode \\
            PPT & LPT  
        \end{tabular}
        \caption{Burn-in of the posterior initial matter power spectra. The left panel corresponds to PPT with annealing, and the right panel corresponds to standard LPT. The colour scale shows the evolution of the matter power spectrum with the number of samples. The dashed lines indicate the underlying power spectrum and the 1- and 2-$\sigma$ cosmic variance limits. The Markov chain is initialised with a Gaussian initial density field scaled by a factor $10^{-3}$ and the amplitudes of the power spectrum systematically drift towards the fiducial values, recovering the true matter power spectrum at the end of the warm-up phase. Monitoring this drift allows us to identify when the Markov chain approaches a stationary distribution and provides unbiased estimates of the target distribution. The annealing with PPT reduces significantly the number of samples required in the warm-up phase, moving the chain faster to the target distribution. This is achieved by first sampling the coarser scales and gradually allowing the algorithm to respond to increasingly finer scales.}
        \label{fig:pk}
    \end{figure*}

    \begin{figure*}[t]
        \centering
            \begin{tabular}{c c}
            \includegraphics[width=0.45\hsize,clip=true]{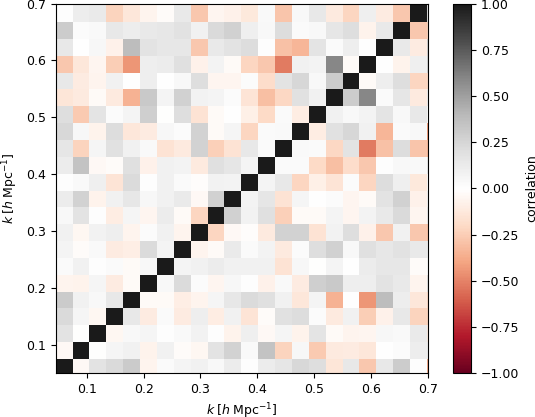} &  
            \includegraphics[width=0.45\hsize,clip=true]{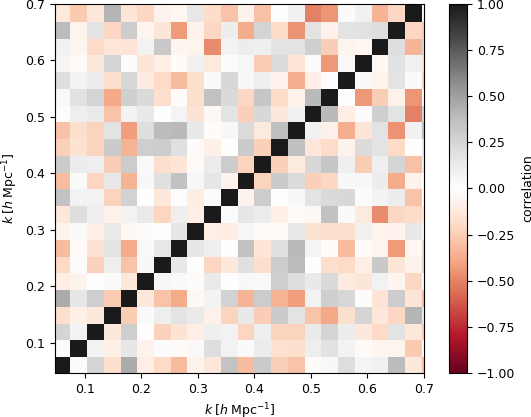} 
            \leavevmode \\
            PPT & LPT
        \end{tabular}
        \caption{Estimated correlation matrix of power spectrum amplitudes with the mean value, normalised using the variance of amplitudes of the power spectrum modes, for PPT (left panel) and LPT (right panel). We computed the correlation matrix from 600 samples after the warm-up phase. The low off-diagonal terms indicate that the annealing method does not introduce any erroneous mode coupling. 
        }\label{fig:pk_autocorr}
    \end{figure*}

    \begin{figure}[t]
        \centering
        \includegraphics[width=\hsize,clip=true]{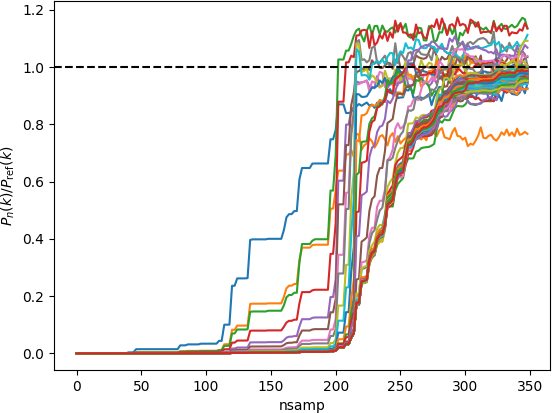}
        \includegraphics[width=\hsize,clip=true]{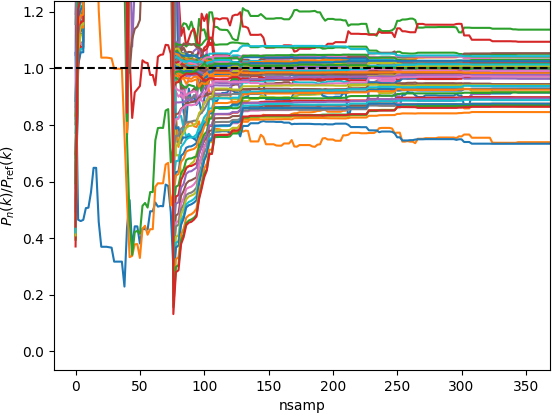}
        \caption{Amplitudes of the posterior primordial matter power-spectrum at different Fourier modes traced during the warm-up phase of the MCMC sampler for the LPT (upper panel) and PPT with annealing (lower panel). As can be seen, initially, modes perform a coherent drift towards the high probability region in posterior distribution and start oscillating around their fiducial values once the Markov chain has reached a stationary state. The fiducial values are reached faster with the PPT annealing, reducing the computational cost of the warm-up phase.
        }\label{fig:burnin}
    \end{figure}
    
    \subsection{Annealing strategy}
    As discussed above, our inference method employs an MCMC sampler. In the large sample limit, any properly set up Markov chain is guaranteed to approach a stationary distribution that provides an unbiased estimate of the target distribution. While Markov chains are typically initialised from a place remote from the target distribution, after a finite amount of transition steps, the chain acquires a stationary state. Once the chain is in the stationary state, we may start recording samples to perform statistical analyses of the inference problem. The initial warm-up phase of the Markov sampler can be costly since it typically requires a high number of samples. In this work, we introduced simulated annealing to reduce the computational cost of the warm-up phase.
    
    The idea behind the simulated annealing is to take advantage of the lower complexity at larger scales and work down a hierarchy from the coarsest to the finest scales of the density field. For that, we start sampling only the large scales of the density field and, once these are converged, we map the density into a finer resolution and sample higher $k$-modes. This process reduces the computational cost of the warm-up phase in two ways. First, we only need to iterate enough to allow the relatively local structure to converge since the larger structures already converged at coarser levels. Secondly, the number of modes to be sampled in the coarser resolution is smaller, allowing rapid sampling.  
    
    When we increase the resolution, a key question is how to anneal in a way that the features of interest are represented in the current level and mapped to the next finer resolution. To answer this question, we make use of insights from re-normalisation theory, as it has been previously done in the field of image processing \citep[see, e.g.][]{Annealing89, Annealing03}. The insight from re-normalisation theory is that the effective temperature for a given feature is scale-dependent. This temperature can be translated into the smoothness of the density field: at some intermediate resolution, coarse scales are converged ('frozen'), and finer scales are still evolving ('hot'). This means that we can concentrate on the intermediate scales and ignore the changes in the smaller scales: we focus on the coarsest not-converged scales. In the PPT model, this is possible by changing the $\hbar$ parameter that controls the effective phase space resolution. We can think of $\hbar$ as an effective temperature, with high $\hbar$ corresponding to high temperature (coarser resolution). Gradually decreasing $\hbar$ corresponds to allowing the algorithm to respond to increasingly finer structures. For high $\hbar$, therefore, we can sample at low resolution since the algorithm only responds to large scales. We then perform a simulated annealing by consistently changing the resolution and $\hbar$. This is illustrated in Fig. \ref{fig:annealing_density}.
    
    In this work, we rely on heuristic rules to determine the changes of $\hbar$ at each level. We start with a high $\hbar$ and a coarse resolution of $N=32^3$ voxels. Once the density field is converged, we open new modes by increasing the number of voxels to $N=64^3$. The algorithm can now respond to these new modes, which are not yet converged. For this reason, we initially keep the same $\hbar$ and reduce it after few iterations, when the evolution of the small scales starts to saturate. Reducing $\hbar$ results in a sharper density and, therefore, the method becomes sensitive to smaller scales. We repeat these steps every time we change the resolution of the density field.

    \subsection{The warm-up phase of the Markov Chain with annealing}
   
    In this Bayesian approach, we keep the cosmology fixed, and specify a prior on the initial power spectrum. However, the power spectrum of the inferred matter distribution is conditioned by the data, and we can use the posterior $P(k)$ as a diagnostic for the effectiveness of the inference since the power spectrum of the simulation differs from the prior. To monitor the initial warm-up phase of the Markov sampler, we follow a similar approach to our previous works \citep{BORG, Foregrounds, Altair, Jasche18, RobustLikelihood, PorqueresLya}: we initialised the Markov chain with an over-dispersed state and traced the systematic drift of inferred quantities towards their preferred regions in the parameter space. Specifically, we initialised the Markov chain with a random Gaussian initial density field scaled by a factor $10^{-3}$ and monitored the drift of corresponding posterior power-spectra during the warm-up phase. Figure \ref{fig:pk} presents the results of this exercise for the standard LPT and the annealing with PPT.  As can be seen, successive measurements of the posterior power-spectrum during the initial warm-up phase show a systematic drift of power-spectrum amplitudes towards their fiducial values. While both forward models correctly recover the fiducial power spectrum, the simulated annealing with the PPT speeds up the burn-in phase, reducing the number of samples needed to reach the target distribution. Fig. \ref{fig:burnin} shows the evolution of the amplitude of the different  Fourier modes in the posterior power spectrum with the number of MCMC samples. While the amplitudes of the $P(k)$ start to evolve in the first 50 samples for the PPT, the modes in the LPT take more than 150 samples to start evolving significantly. These results show that the annealing with PPT allows moving the chain faster towards the high probability regions of the parameter space, reducing the computational cost of the warm-up phase. 
    
    To test for residual correlations between different Fourier modes, we estimated the covariance matrix of power-spectrum amplitudes from our ensemble of Markov samples. Figure \ref{fig:pk_autocorr} shows that the covariance matrix for PPT with annealing has a clear diagonal structure, equivalent to the covariance for the standard LPT. This confirms that the annealing does not introduce spurious correlations between scales. 
    
    \subsection{Correlation length}
    
    By design, subsequent samples in Markov chains are correlated. The statistical efficiency of an MCMC algorithm is determined by the effective number of independent samples that can be drawn from a chain of a given length. To estimate the statistical efficiency of the sampler, we estimate the correlation length of the density amplitude at different locations of the box. For the amplitude at a given voxel, $\theta$, the auto-correlation for samples with a given lag in the chain can be estimated as
    \begin{equation}
        C_n(\theta) = \frac{1}{N-n} \sum_{i=0}^{N-n} \frac{(\theta^i - \langle \theta\rangle)(\theta^{i+n} - \langle \theta \rangle)}{\mathrm{Var}(\theta)}
    \end{equation}
    where $n$ is the lag in MCMC samples, $\langle \theta\rangle$ is the mean and $\mathrm{Var}(\theta)$ is the variance. We typically determine the correlation length by estimating the lag $n_C$ at which the auto-correlation $C_n$ dropped below $0.1$. The number $n_C$ therefore presents the number of transitions required to generate one more effectively independent sample. 
    
\begin{figure}[t]
        \centering
        \includegraphics[width=\hsize,clip=true]{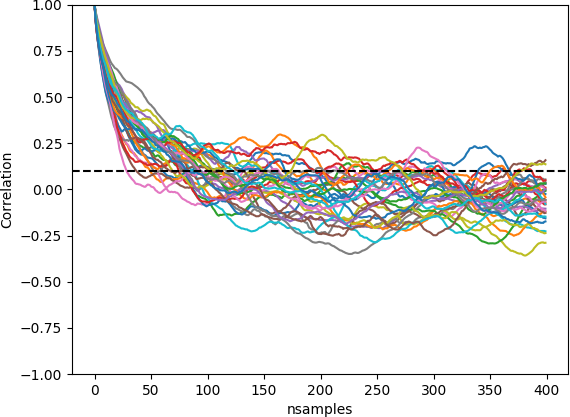}
        \includegraphics[width=\hsize,clip=true]{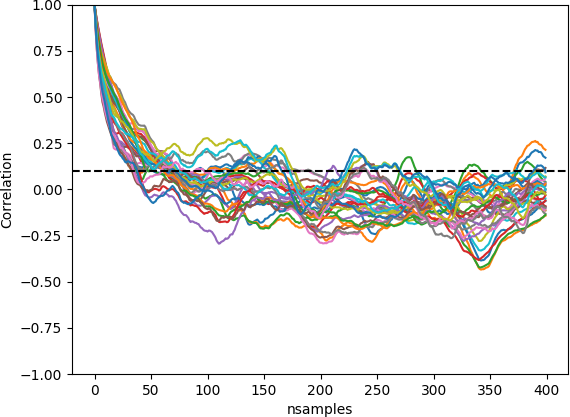}
        \caption{Autocorrelation of the density amplitudes as a function of the sample lag in the Markov chain for LPT (upper panel) and PPT (lower panel). This autocorrelation is estimated from 2000 samples after the warm-up phase. The correlation length of the sampler can be estimated by determining the point when correlations drop below 0.1 for the first time. The annealing with the PPT model does not introduce longer correlations in the density sampler.
        }\label{fig:correlation}
    \end{figure}

    Fig. \ref{fig:correlation} presents the results of this test for the standard LPT and the annealing with PPT. As can be seen, the correlation is generally lower for the PPT. This confirms that reducing the computational costs of the warm-up phase by annealing does not come at the expense of introducing longer correlations in the chain. Therefore, the annealing with the PPT results in a net speed-up of the Markov sampler to reach the target distribution. 
    
    Without annealing, the PPT shows an equivalent warm-up phase and correlation length to LPT (350 samples). However, the PPT is still an advantage over LPT since it provides a more accurate description of the density at low-density regimes (see Section \ref{sec:comparison_lpt})

    \subsection{Comparison to LPT} \label{sec:comparison_lpt}
    
    In this section, we compare the large-scale structures obtained with LPT and PPT. More specifically, this section focuses on comparing the profiles of cosmic structures in the density fields. For this, we evolved a set of initial conditions with both forward models (PPT and LPT) and compared the profiles of individual voids and clusters.
    
\begin{figure}[t]
        \centering
            {\includegraphics[width=\hsize,clip=true]{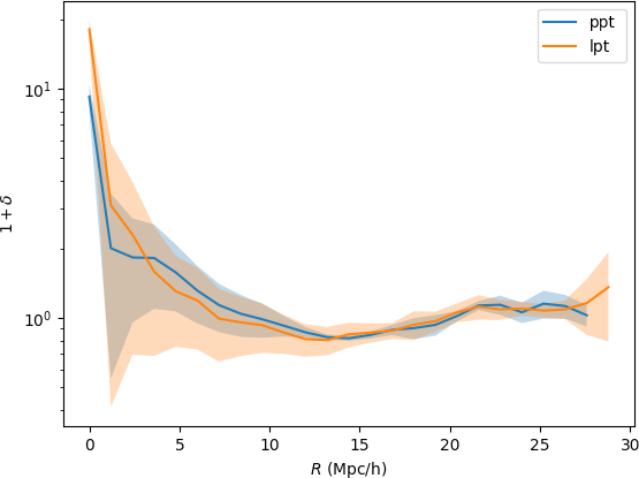}}
            {\includegraphics[width=\hsize,clip=true]{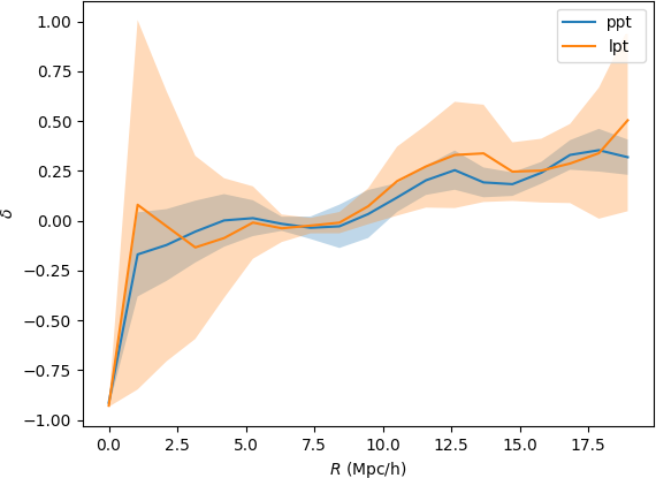}}
        \caption{Comparison of density profiles obtained with PPT and LPT. The upper panel shows a cluster density profile obtained by evolving the same initial conditions with PPT and LPT. The lower panel shows the test for a void profile. This demonstrates that PPT and LPT provide equivalent cosmic structures. The shaded regions indicate the uncertainty region of the profiles, corresponding to the standard deviation of 50 realisations. While the PPT and LPT show similar uncertainty regions for the cluster profile, the LPT has a larger standard deviation than the PPT in the void profile. This larger uncertainty is due to the poor sampling of the voids in the LPT model since most of the particles cluster in high-density regions.}
        \label{fig:profiles}
    \end{figure}

    To compare the density profiles of a cluster (or a void), we randomly chose a local maximum (or minimum) in the final density field. We, then, determined the density profiles in spherical concentric shells. Figure \ref{fig:profiles} shows the density profiles for a cluster and a void obtained with PPT and LPT. Both models provide the same profiles, indicating that the PPT and LPT describe equivalent cosmic structures. While the standard deviation of the cluster profile is similar for both methods, the void profile shows a larger uncertainty region for LPT. This larger uncertainty in the void is due to the particle sampling noise introduced by the LPT. Since most of the particles cluster in high-density regions, voids are impacted by higher uncertainty in the LPT. The field-level approach of the PPT overcomes this problem, showing a lower standard deviation in the void profile.
    
\section{Inference results}
    
    \begin{figure*}
        \centering
        {\includegraphics[width=\hsize,clip=true]{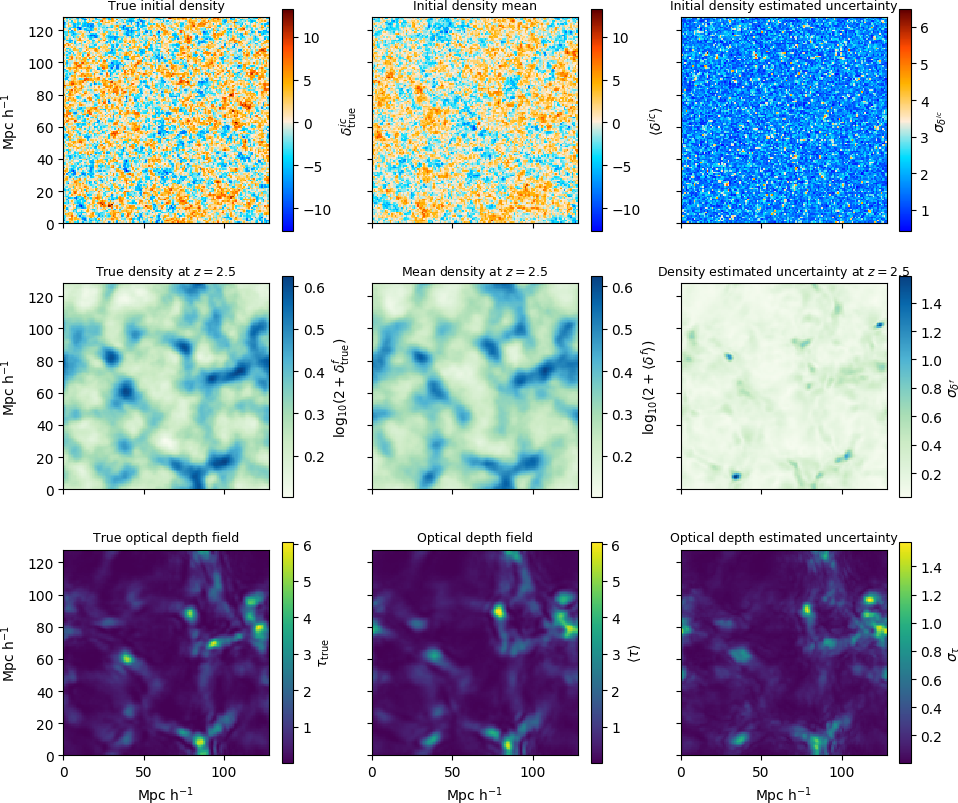}}
        \caption{Slices through ground truth initial (left upper panel), true evolved density field (left middle panel), true optical depth field (left lower panel), inferred ensemble mean initial (middle upper panel), ensemble mean evolved (middle-lower panel) density field and ensemble mean optical depth (lower middle panel) computed from 600 MCMC samples. The density fields are in physical space, obtained with the PPT as indicated in eq. \ref{eq:rho}. The optical depth field is in redshift space, corresponding to $\tau=\chi \bar{\chi}$ with $\chi$ from eq. \ref{eq:chi}. Comparison between these panels shows that the method recovers the structure of the true density fields with high accuracy. Right panels show standard deviations of inferred amplitudes of initial (upper right panel), final density fields (middle right panel) and optical depth (lower right panel). We note that we plotted the standard deviation of the density $\sigma_{\delta^f}$ but the mean density is plotted as $\log_{10}(2+\langle \delta^f \rangle)$. We note that the uncertainty of $\delta^\mathrm{f}$ and of the optical depth present a structure that correlates with the corresponding field. In contrast, the standard deviation of the initial conditions are homogeneous and show no correlation with the initial density field, indicating that the dynamical model correctly propagates the information between the primordial matter fluctuations and the final density and absorption fields. } 
        \label{fig:panels}
    \end{figure*}

     \begin{figure}
        \centering
        {\includegraphics[width=\hsize,clip=true]{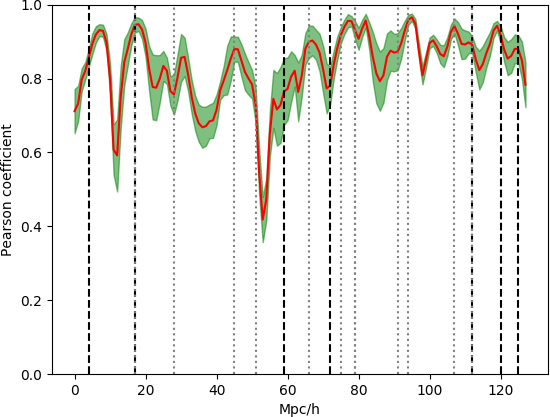}}
        \caption{Pearson coefficient of the true density field and 300 density samples in our Markov chain. The red line corresponds to the mean of the correlation, and the shaded region indicates the standard deviation. The correlation is computed for a slice of width 1 Mpc/h (the x-axis indicate the position across this slice). The dashed lines indicate the position of the lines of sight in the slice, and the dotted lines indicate the position of lines of sight in neighbouring slices. The Pearson coefficient is > 0.7 at most of the locations in this slice, including regions where there are no neighbouring lines of sight. This indicates that the algorithm can interpolate the information between lines of sight and correctly recover the structures in unobserved regions.}
        \label{fig:pearson}
    \end{figure}
    
    \begin{figure}[t]
        \centering
         {\includegraphics[width=\hsize,clip=true]{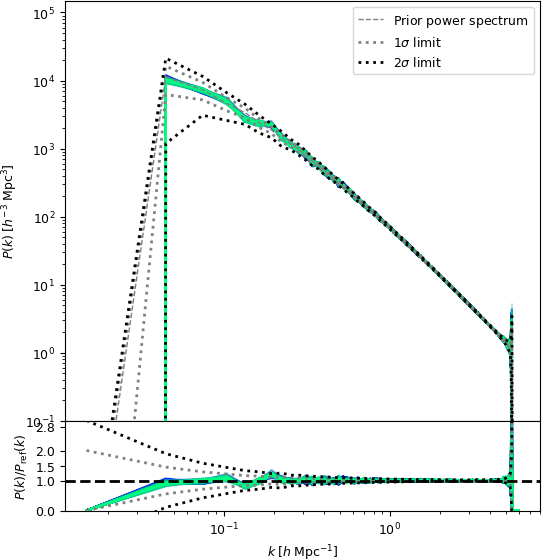}}
         \caption{Mean posterior matter power-spectrum. The mean and the standard deviation of the initial matter power spectrum have been computed from 300 density samples of the Markov chain obtained after the warm-up phase. The standard deviation is plotted, but it is too small to be visible, showing the stability of the posterior power-spectrum. The dashed line indicates the underlying power spectrum and the 1- and 2-$\sigma$ cosmic variance limit. The algorithm recovers the fiducial power-spectrum amplitudes within the 1-$\sigma$ cosmic variance uncertainty limit throughout the entire range of Fourier modes considered in this work.
        \label{fig:Pk_rsd}}
    \end{figure}
    
    In this section, we present the results of applying our algorithm to \lya forest data in redshift space. We show that our method infers unbiased density fields and corresponding power-spectra at all scales considered in this work. We also perform a posterior predictive test for quasar spectra, showing that the inferred quantities can explain the data within the noise uncertainty. 
    
    \subsection{Inferred density fields} \label{sec:run_qlpt}
    
    As discussed above, our method uses a forward modelling approach, fitting a physical dynamical model to \lya forest data. This provides the full posterior distribution, from which we draw samples of the initial matter fluctuations and the non-linear spatial matter distribution at $z=2.5$. In this section, the dynamical model is the PPT in redshift space, described in Section \ref{sec:qlpt_rsd_lya}. Since the optical depth is conserved under the mapping between physical and redshift space, our inference in redshift space focuses on the optical depth field, while the density field is obtained in real space.  
    
    Figure \ref{fig:panels}  shows slices through the true fields, and the ensemble mean and variances of inferred three-dimensional fields, computed from 600 samples. A first visual comparison between ground truth and the inferred ensemble mean final density and optical depth fields shows that the algorithm correctly recovered the large-scale structure from Ly-$\alpha$ forest data. We note that the optical depth field is in redshift space while the final density field is in physical space. The lower right panel of figure \ref{fig:panels} shows the corresponding standard deviations of the amplitudes, which is estimated from the samples in the Markov Chain. The estimated density standard deviation correlates with the inferred density field. The same is true for the optical depth field. This is expected for a non-linear data model, which couples signal and noise. Higher uncertainty regions correspond to over-densities since the absorption saturates at high density, making the signal weaker. Once the light absorption is saturated, the data only provide information on a minimally lowest density threshold required to explain the observations. The line saturation effectively removes constraints from data above some lower threshold, nullifying the impact of higher density amplitudes in the likelihood. This leads the algorithm to use solely the prior to fill the missing pieces in high-density regions. The inference is thus not impacted at all by this observational limitation.
    In future development of the method, tighter constraints of the density amplitude at high densities could be achieved by modelling the absorption line profile since the line saturation introduces broadening of the profile.
    
    While the standard deviations of the fields at $z=2.5$ present a structure that correlates with the density, the standard deviation of the initial conditions is Gaussian noise, as shown in the upper panels of Fig. \ref{fig:panels}. This indicates that our forward model correctly propagates the information between the initial and final density field.
    
    As discussed above, the mean separation between lines of sight is the most relevant parameter in \lya forest surveys. A particular challenge is to recover the density field in between one-dimensional lines of sight. To test the performance of our algorithm, we computed the Pearson correlation coefficients between the true density field and several density samples in our Markov chain. Figure \ref{fig:pearson} shows the correlation coefficients for a slice of width 1 Mpc/h, indicating the value of the Pearson coefficient at different locations across this slice. This permits us to track the correlations on and in-between lines of sight (indicated with dashed lines). At the position of the lines of sight, the correlation is typically > 80 \%. For a better understanding of the fluctuations in the correlation, we indicated the position of lines of sight in the neighbouring slices (dotted lines). Most of the dotted lines correspond to a peak in the correlation, which indicates that the algorithm can interpolate the information between lines of sight. Also, there are regions of 15 Mpc/h without neighbouring lines of sight where the correlation is still >70 \%  (see the regions centred at 36 and 100 Mpc/h), indicating that the method can recover the cosmic large-scale structure in the unobserved regions between observed lines of sight. We note that this plot is not comparable to Fig. 11 in \cite{PorqueresLya} since the signal-to-noise of the mock data, the distribution of lines of sight and the width of the slice are different. A more detailed analysis of the impact of the mean line of sight separation on the accuracy of the results will be included in future work.
    
    As a posterior test, we estimate the mean and variance of posterior power-spectra measured from the ensemble of Markov samples. The result is shown in Fig. \ref{fig:Pk_rsd}. Although our method does not sample the different modes of the power spectrum, this is not enforced on the density samples. This means that, if the data requires it, the primordial power spectrum can be overwritten. Reconstructing three-dimensional density fields from one-dimensional Ly-$\alpha$ data is technically challenging. Previous approaches of inferring the density field from the \lya forest failed at recovering the correct power-spectrum amplitudes. For example, \cite{Kitaura12} used a Gibbs sampling approach to sample the large- and small-scales of the density field separately with a log-normal prior for the evolved density field. This approach inferred correct power-spectrum amplitudes at large scales $k<0.1$ h Mpc$^{-1}$ but obtained erroneous excess power at smaller scales. \cite{Horowitz19} used an optimisation approach to fit a dynamical forward model to the data, but the method obtains power-spectra that severely underestimate the power of density amplitudes. Typically, deviations from the fiducial power spectrum indicate the breakdown of the assumptions in the data model or the inference method. Figure \ref{fig:Pk_rsd} shows that our method recovers the fiducial power spectrum within the 1-$\sigma$ cosmic variance uncertainty at scales considered in this work. This demonstrates that our method is capable of inferring the matter distributions with the correct power spectrum from noisy Ly-$\alpha$ data in redshift space.

    \begin{figure}[t]
        \centering
         {\includegraphics[width=\hsize,clip=true]{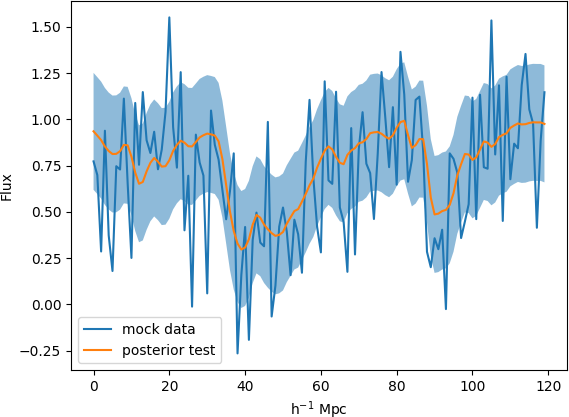}}
        \caption{Posterior predictive flux for a spectrum with noise $\sigma = 0.03$. The posterior predicted flux (orange line) is computed from the ensemble mean optical depth field in redshift space. The blue shaded region indicates the 1-$\sigma$ region, corresponding to the standard deviation of the noise in this line of sight. This test checks whether the data model can accurately account for the observations. Any significant mismatch would immediately indicate a breakdown of the applicability of the data model or error of the inference framework. Our method recovers the transmitted flux fraction correctly within the noise uncertainty. 
        \label{fig:posterior_test}}
    \end{figure}
    
    \begin{figure}[t]
        \centering
            {\includegraphics[width=\hsize,clip=true]{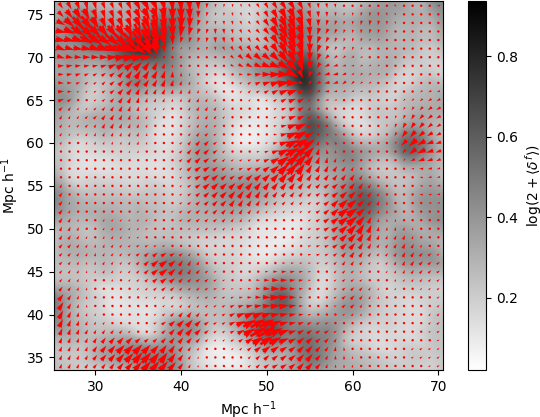}}
        \caption{Zoom-in on the density field. The vector field shows the velocities derived from PPT, showing matter flowing out of the void and falling into the gravitational potential of the cluster. Our method provides consistent velocity and density fields that can be used to study structure formation. In particular, the velocity derived from the PPT provides more accurate estimates than the standard LPT in voids and filaments.}
        \label{fig:velocity}
    \end{figure}
    
    \begin{figure*}
        \centering
        {\includegraphics[width=\hsize,clip=true]{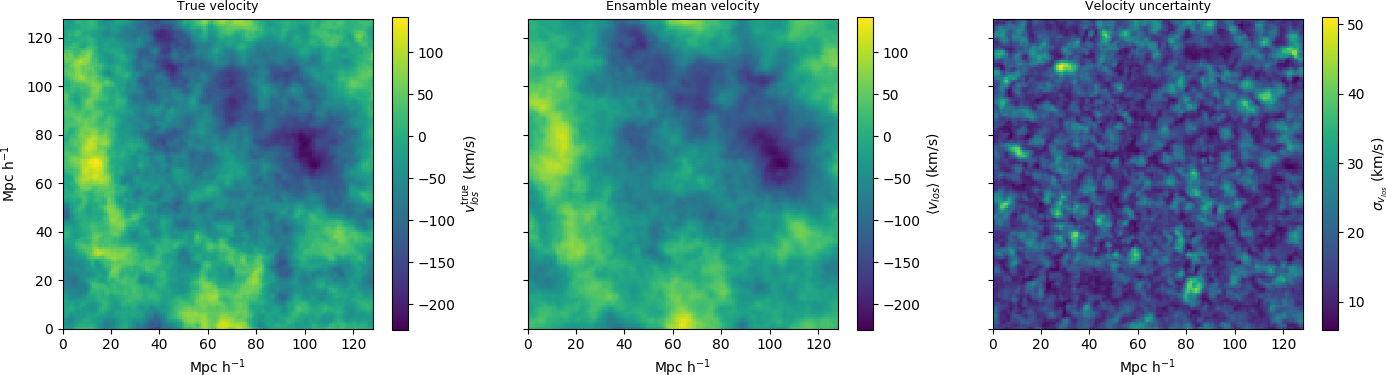}}
        \caption{Slices through ground truth (left panel) and mean (middle panel) velocity field in the direction of the line of sight, estimated from 200 samples. Comparison between these panels shows that the method recovers the true velocity field. The right panel shows the standard deviation.} 
        \label{fig:vel_panels}
    \end{figure*}
    
    \subsection{Posterior predictive tests}
    
    Posterior predictions allow testing of whether the inferred density fields provide accurate explanations for the data \citep[see, e.g.][]{gelmanbda04}. Generally, posterior predictive tests provide good diagnostics about the adequacy of data models in explaining observations and identifying possible systematic problems with the inference. Figure \ref{fig:posterior_test} shows the result of this test for one line of sight in redshift space, showing that the posterior predicted quasar spectrum recovers the data input within the observational $1\sigma$ uncertainty region. The $1\sigma$ region corresponds to the standard deviation of the noise added in this line of sight. This demonstrates that the method correctly locates absorber positions and corresponding amplitudes of the underlying densities and, therefore, the inferred quantities can explain the data at the level of the noise.
    
    \subsection{Velocity fields from the PPT}
    
    The dynamical model in our algorithm allows us to naturally infer the velocity field since it derives from the initial perturbations. This velocity information can provide significant information on the formation of structures since it allows discrimination between peculiar velocities and the Hubble flow. 
    
    Figure \ref{fig:vel_panels} shows a slice through the line-of-sight component of the velocity field from the ground truth and the mean and standard deviation estimated from 200 samples. A visual comparison between the true and mean velocity fields shows that the algorithm recovers the true velocity from \lya forest data. This method, therefore, provides velocity fields constrained by the data at $z>2$, where this information is challenging to obtain otherwise.
    
    Figure \ref{fig:velocity} shows a zoom-in on the inferred mean density field and the corresponding velocity components $(v_x,v_y)$ obtained with PPT. PPT provides more accurate peculiar velocities than the LPT in voids and filaments. In the LPT approach, one obtains the momentum field since the velocities are associated with massive particles. This means that we need to divide by the density field to obtain the peculiar velocities $\vec{v} = \vec{j} / \rho$. This requires smoothing of the density field to avoid empty regions with $\rho=0$ and, in filaments, we need to average over enough particles. PPT overcomes these problems by operating on the field.

\section{Summary and discussion} \label{sec:conclusions}

    While LPT provides a good description of the dynamics of cold dark matter before shell-crossing, it requires to interpolate the fluid elements to an Eulerian field to obtain density and velocity fields in physical space. The discrete fluid elements further introduce a particle sampling noise in the fields. Since most of the particles cluster in the over-densities, voids and under-dense regions are more impacted by the sampling noise. These under-dense regions are especially relevant when analysing \lya forest observations since these data mostly arise from sheets and voids.
    
    A recent alternative to LPT is a field-based approach that directly predicts the Eulerian density field. The PPT, presented in \cite{Ulhemann19}, provides such an alternative to LPT, overcoming particle sampling noise, and giving easy access to the full Boltzmann hierarchy. PPT uses a propagator to evolve a wave function that encodes the density and velocity (and higher moment) information. 
    
    In this work we employ, for the first time, PPT in a Bayesian forward model to infer primordial fluctuations from sparse redshift-space quasar flux spectra, connected by the dynamical model and the fluctuating Gunn-Peterson approximation (FGPA). This framework is based on a Gaussian prior for the primordial fluctuations and a likelihood based on the quasar flux field for the \lya forest. We fixed the cosmology. To explore the parameter space, our method employs MCMC techniques.

    Furthermore, the PPT approach introduces a free parameter, $\hbar$, that acts as a natural smoothing (or temperature) scale. Allowing $\hbar$ to evolve over time allows performing a simulated annealing, thereby selecting the features and scales that the algorithm is sensitive to. Taking advantage of the lower complexity of coarser scales, we decrease $\hbar$ over the course of the chain, allowing the algorithm to respond to increasingly finer structures. By comparing to the standard LPT, we have shown that the PPT annealing reduces the computational cost of the warm-up phase of the MCMC sampler. With our implementation serving as a proof-of-concept, we find that multi-scale techniques are able to accelerate MCMC burn-in significantly. More sophisticated algorithms might be possible in the future that exploit this aspect further.
    
    Since cosmological observations take place in redshift space, we have derived the RSD within the PPT formalism. We showed that RSDs can be easily included via an additional propagator between physical Eulerian space and redshift space. Since the optical depth is conserved under the mapping between physical and redshift space, the RSD propagator can be applied to a wave function that directly encodes the optical depth instead of the density. Based on PPT, we are therefore able to provide a forward model mapping primordial fluctuations to quasar flux in redshift space at the field level.
    
    We have tested our inference method in redshift space with simulated data. These tests showed that our method recovers the underlying initial and final density field (in physical space) and the optical depth field (in redshift space). Our method based on PPT is able to correctly propagate the information between the initial matter fluctuations and the density and absorption field at $z=2.5$. 
    
    From the dynamical forward model, we can easily derive peculiar velocity fields constrained by the data. To obtain peculiar velocities with the standard LPT, one needs to divide the momentum field by the density. This requires to previously smooth the density field to avoid empty regions and, in filaments, one needs to average over a sufficient number of particles to obtain the correct velocity in the filament. These problems are naturally overcome by PPT since it operates at the field level. Therefore, PPT provides a better estimate of peculiar velocities in filaments and voids.  This clearly demonstrates the advantage of field-based over fluid-element based dynamical models.
    
\section*{Acknowledgements}
    NP and OH thank Alan Heavens, Cornelius Rampf and Cora Uhlemann for discussions and comments on the draft. NP acknowledges funding from STFC through Imperial College Astrophysics Consolidated Grant ST/5000372/1. OH acknowledges funding from the European Research Council (ERC) under the European Union’s Horizon 2020 research and innovation programme (Grant Agreement No. 679145, project “COSMO-SIMS”).  GL acknowledges financial support from the ILP LABEX, under reference ANR-10-LABX-63, which is financed by French state funds managed by the ANR within the programme `Investissements d'Avenir' under reference ANR-11-IDEX-0004-02. GL also acknowledges financial support from the ANR BIG4, under reference ANR-16-CE23-0002. This work was carried out within the Aquila Consortium\footnote{\url{https://aquila-consortium.org}}.

\bibliographystyle{aa} 
\bibliography{biblio} 

\appendix

\section{Adjoint gradient of the PPT} \label{app:qlpt_gradient}
    The inference of the density field requires inferring the amplitudes of the primordial density at different volume elements of a regular grid, commonly between $128^3$ and $256^3$ volume elements. This implies $10^6$ to $10^7$ free parameters. To explore this high-dimensional parameter space efficiently, the BORG framework employs a Hamiltonian Monte Carlo (HMC) method, which adapts to the geometry of the problem by using the information in the gradients. Therefore, this algorithm requires the derivatives of the forward model. In this section, we derive the gradient of the PPT. 

    More specifically, the HMC relies on the availability of a gradient of the posterior distribution. Therefore, we need to compute the gradient of the log-likelihood with respect to the gravitational potential $\phi^\mathrm{ic}$. 
    
    \subsection{Gradient of PPT in real space}
    First, we derive the gradient corresponding to the PPT described in Section \ref{sec:QLPT_background} and \ref{sec:implementation}. The gradient of PPT in redshift space is derived in the following section. 
    
    We want the gradient of the likelihood with respect to the gravitational potential $\phi^\mathrm{ic}_p$,
    \begin{equation}
       \frac{\partial \log\mathcal{L}} {\partial \phi^\mathrm{ic}_p} = \sum_l \frac{\partial \log\mathcal{L}}{\partial \delta^\mathrm{f}_l} \frac{\partial \delta^\mathrm{f}_l}{\partial \phi^\mathrm{ic}_p}
       \label{eq:A1}
    \end{equation}
    where $\log\mathcal{L}$ is the log-likelihood function and $\delta^\mathrm{f}_l$ indicates the final density field at voxel $l$. In the Gunn-Peterson approximation, the derivative of the data model is
    \begin{eqnarray}
        \frac{\partial \log\mathcal{L}}{\partial \delta^\mathrm{f}_l} &=& \sum_n \frac{(F_n)_l - \exp[-A(1+\delta^\mathrm{f})_l)^{\beta}]}{\sigma^2} \nonumber \\ 
&\times& A \beta \big(1+\delta^\mathrm{f})_l\big)^{\beta-1} \exp\Big[-A\big(1+\delta^\mathrm{f})_l\big)^{\beta}\Big]
    \end{eqnarray}
    where $n$ runs over the different lines of sight. However, in this section we focus on the gradient of the dynamical forward model, indepently of the data model. Therefore, we not specify the derivative of the likelihood with respect to the final density. 
    
    The final density field is given by $\delta^\mathrm{f}_l=\psi_l \bar{\psi}_l$. Therefore, eq. \ref{eq:A1} reads
    \begin{equation}
        \frac{\partial \log\mathcal{L}} {\partial \phi^\mathrm{ic}_p} 
        = \sum_l \frac{\partial \log\mathcal{L}}{\partial \delta^\mathrm{f}_l} \left[ \frac{\partial \psi_l}{\partial \phi^\mathrm{ic}_p}\bar{\psi}_l + \psi_l \frac{\partial \bar{\psi}_l}{\partial \phi^\mathrm{ic}_p} \right].
    \end{equation}
    
    The derivative of $\psi$ with respect to the initial gravitational potential is 
    \begin{equation}
        \frac{\partial \psi_l}{\partial \phi^\mathrm{ic}_p} = \sum_m M_{lm} \hat{K}_m M'_{mp} \left(\frac{-i}{\hbar}\right) e^{-\frac{i}{\hbar}\phi_p^\mathrm{ic}}
    \label{eq:der_psi}
    \end{equation}
    and 
    \begin{equation}
        \frac{\partial \bar{\psi}_l}{\partial \phi^\mathrm{ic}_p} = \sum_m M_{lm} \overline{\hat{K}}_m M'_{mp} \left(\frac{i}{\hbar}\right) e^{\frac{i}{\hbar}\phi_p^\mathrm{ic}}
    \label{eq:der_psi_hat}
    \end{equation}
    where $M$ indicate Fourier transforms and $\hat{K}_m = \exp \left[-i\frac{1}{2}\hbar k_m^2 D_+\right]$ is the free propagator in Fourier space. 
    
    Finally, the gradient reads
    \begin{eqnarray}
    \frac{\partial \log\mathcal{L}} {\partial \phi^\mathrm{ic}_p} &=&
    \frac{i}{\hbar} \sum_l \frac{\partial \log\mathcal{L}}{\partial \delta^\mathrm{f}_l}   \\ \nonumber 
    &\times& \sum_{m} M_{lm} \left[ \overline{\hat{K}}_m M'_{mp}  e^{\frac{i}{\hbar} \phi_p}
    - \hat{K}_m M'_{mp} e^{\frac{-i}{\hbar} \phi_p} \right] .
    \end{eqnarray}
    
    \subsection{Gradient PPT with RSD}
    Here we derived the gradient of the PPT in redshift for the Ly-$\alpha$ forest, described in Section \ref{sec:qlpt_rsd_lya}. As before, the HMC requires the gradient of the likelihood with respect to the gravitational potential $\phi^{ic}$. 
    \begin{equation}
       \frac{\partial \log\mathcal{L}} {\partial \phi^\mathrm{ic}_p} = \sum_l \frac{\partial \log\mathcal{L}}{\partial \tau_l} \frac{\partial \tau_l}{\partial \phi^\mathrm{ic}_p}
    \label{eq:A2}
    \end{equation}
    where $\tau_l$ is the optical depth at voxel $l$. In this case, the derivative of the data model is 
    \begin{eqnarray}
        \frac{\partial \log\mathcal{L}}{\partial \tau_l} &=& - \sum_n \frac{(F_n)_l - \exp(-\tau_l)}{\sigma^2}  \exp(-\tau_l)
    \end{eqnarray}
    where $n$ runs over the lines of sight. 
    
    The optical depth is given by $\tau = \chi \bar{\chi}$. Therefore, eq. \ref{eq:A2} is
    \begin{equation}
        \frac{\partial \log\mathcal{L}} {\partial \phi^\mathrm{ic}_p} 
        = \sum_l \frac{\partial \log\mathcal{L}}{\partial \tau_l} \left[ \frac{\partial \chi_l}{\partial \phi^\mathrm{ic}_p}\bar{\chi}_l + \chi_l \frac{\partial \bar{\chi}_l}{\partial \phi^\mathrm{ic}_p} \right].
    \label{eq:2terms_rsd}
    \end{equation}
    Developing the first term, we obtain
    \begin{eqnarray}
        \sum_l \frac{\partial \log\mathcal{L}}{\partial \tau_l}  \frac{\partial \chi_l}{\partial \phi^\mathrm{ic}_p}\bar{\chi}_l &=&   \sum_l \frac{\partial \log\mathcal{L}}{\partial \tau_l} \bar{\chi_l} \\ \nonumber  &\times& \sum_m M_{lm} \left(\hat{K}_\mathrm{RSD}\right)_m \sum_n M'_{mn} \frac{\partial \chi_0}{\partial \phi^\mathrm{ic}}
    \end{eqnarray}
    where $M$ indicate a Fourier transform and $\hat{K}_\mathrm{RSD}$ is the propagator in eq. \ref{eq:propagator_rsd}. 

    Introducing $\chi_0 = \sqrt{A} \left(\psi\bar{\psi}\right)^{\frac{\beta-1}{2}} \psi$ in the previous equation, we get
    \begin{eqnarray}
        \sum_l \frac{\partial \log\mathcal{L}}{\partial \tau_l}  \frac{\partial \chi_l}{\partial \phi^\mathrm{ic}_p}\bar{\chi}_l &=&   \sqrt{A} \sum_l \frac{\partial \log\mathcal{L}}{\partial \tau_l} \bar{\chi_l} \\ \nonumber  &\times& \sum_m M_{lm} \left(\hat{K}_\mathrm{RSD}\right)_m \\ \nonumber
        &\times& \sum_n M'_{mn} \left[ \left(\psi\bar{\psi}\right)_n^{\frac{\beta-1}{2}}   \frac{\partial \psi_n}{\partial \phi^\mathrm{ic}}\right.  \\ \nonumber 
        &+& \left. \frac{\beta-1}{2} \left(\psi\bar{\psi}\right)_n^{\frac{\beta-3}{2}} \left( \frac{\partial \psi_n}{\partial \phi^\mathrm{ic}_p}\bar{\psi}_n + \psi_n \frac{\partial \bar{\psi}_n}{\partial \phi^\mathrm{ic}_n} \right) \psi_n \right] .
    \end{eqnarray}
    
    Similarly, we can obtain the second term of eq. \ref{eq:2terms_rsd} and combine them, obtaining
    \begin{eqnarray}
        \sum_l \frac{\partial \log\mathcal{L}}{\partial \tau_l}  \frac{\partial \chi_l}{\partial \phi^\mathrm{ic}_p}\bar{\chi}_l &=&   \sqrt{A} \sum_l \frac{\partial \log\mathcal{L}}{\partial \tau_l} \bar{\chi_l} \\ \nonumber  
        &\times& \sum_m M_{lm} \Bigg\{\Bigg[ \left(\hat{K}_\mathrm{RSD}\right)_m \sum_n M'_{mn} \frac{\beta+1}{2} \rho_n^{\frac{\beta-1}{2}} \\ \nonumber 
        &+& \overline{\left(\hat{K}_\mathrm{RSD}\right)}_m \sum_n M'_{mn} \frac{\beta-1}{2}\rho_n^{\frac{\beta-3}{2}} (\bar{\psi})^2_n \Bigg] \frac{\partial \psi_n}{\partial \phi^\mathrm{ic}_p} \\ \nonumber
        &+& \Bigg[ \left(\overline{\hat{K}_\mathrm{RSD}}\right)_m \sum_n M'_{mn} \frac{\beta+1}{2} \rho_n^{\frac{\beta-1}{2}} \\ \nonumber 
        &+& \left(\hat{K}_\mathrm{RSD}\right)_m \sum_n M'_{mn} 
        \frac{\beta-1}{2}\rho_n^{\frac{\beta-3}{2}} (\psi)^2_n \Bigg]  \frac{\partial \bar{\psi}_n}{\partial \phi^\mathrm{ic}_p} \Bigg\} 
    \end{eqnarray}
    with the derivatives from eq. \ref{eq:der_psi} and \ref{eq:der_psi_hat}.

\end{document}